[1]Mehmet Karahan

# Modeling and Simulation of an Active Car Suspension with a Robust LQR Controller under Road Disturbance, Parameter Uncertainty and White Noise

***Abstract: -*** Vehicle suspension is important for passengers to travel comfortably and to be less exposed to effects such as vibration and shock. A good suspension system increases the road holding of vehicles, allows them to take turns safely, and reduces the risk of traffic accidents. A passive suspension system is the most widely used suspension system in vehicles due to its simple structure and low cost. Passive suspension systems do not have an actuator and therefore do not have a controller. Active suspension systems have an actuator and a controller. Although their structures are more complex and costly, they are safer. The Proportional-Integral-Derivative (PID) controller is widely used in active suspension systems due to its simple structure, reasonable cost, and easy adjustment of coefficients. In this study, a more robust Linear Quadratic Regulator (LQR)-controlled active suspension was designed than a passive suspension and a PID-controlled active suspension. Robustness analyses were performed for passive suspension, PID-controlled active suspension, and LQR-controlled active suspension. Suspension travel, sprung mass acceleration, and sprung mass motion simulations were performed for all three suspensions under road disturbance, under simultaneous road disturbance and parameter uncertainty and under road disturbance with white noise. A comparative analysis was performed by obtaining the rise time, overshoot, and settling time data of the suspensions under different conditions. It was observed that the LQR-controlled active suspension showed the fastest rise time, the least overshoot and had the shortest settling time. In this case, it was proven that the LQR-controlled active suspension provided a more comfortable and safe ride compared to the other two suspension systems.

## I. INTRODUCTION

The suspension system acts as a bridge between the vehicle's chassis and wheels [1,2]. It plays an important role in the vehicle's roadholding, comfort, and safety [3]. It provides vehicle balance and improves braking distance [4]. It increases the comfort level of the driver and passengers by absorbing vibrations and impacts on the road surface [5]. A well-functioning suspension system increases road safety by ensuring that the wheels grip the ground better and reducing the risk of traffic accidents [6]. If the maintenance of the suspension system is neglected, the load on the tires increases, the tires wear out faster, the suspension parts deteriorate, other components are damaged and the need for repairs arises [7]. When the suspension system does not work properly, the ability of the wheels to contact the ground properly decreases [8]. This negatively affects the vehicle's roadholding and driving performance. It increases the vehicle's coefficient of friction and causes more power to be consumed [9]. Therefore, the vehicle's fuel efficiency decreases and fuel consumption increases [10].

Suspension systems are divided into three categories: passive, semi-active, and active [11]. A passive suspension system is a suspension in which the properties of components, such as springs and shock absorbers, are fixed. The mechanical structure of passive suspension systems is simple and low-cost. For this reason, it is widely preferred [12]. A semi-active suspension system can adjust the stiffness of the suspension according to changing road conditions [13]. In semi-active suspension, electro-rheological (ER) and magneto-rheological (MR) shock absorbers are used instead of classic springs and shock absorbers [14]. ER and MR shock absorbers can adjust the stiffness and damping coefficients according to different conditions by adjusting the electric current and magnetic field [15]. The active suspension system, unlike the semi-active suspension, applies an extra actuator between the vehicle body and the wheel axle to add and distribute energy from the system [16]. Thus, an improved suspension response is achieved compared to passive and semi-active suspensions. Although the active suspension system requires higher energy consumption and a complex mechanical structure, it has much better performance than other suspension systems [17].

Kumar et al. used a half-truck model and developed a passive suspension for this model. They optimized their passive suspension with a genetic algorithm. They aimed to minimize road damage with this algorithm. They made their modeling and simulations with MATLAB and Simulink [18]. Issa et al. proposed a passive suspension optimized with the Harris Hawk Optimization algorithm. They tested this algorithm on quarter and semi-car models. They compared the algorithm they proposed with Particle Swarm Optimization, Genetic Algorithm, and the Fire-Fly Optimization algorithm. The model they proposed reduced body acceleration by 16.5% compared to the classical passive suspension. They used MATLAB in their work. [19]. Lee et al. developed a semi-active suspension for a quarter car model. They used a deep reinforcement learning algorithm for optimization. They compared the performance of the on-policy reinforcement algorithm and the off-policy reinforcement algorithm for their semi-active suspension [20]. Ab Talib et al. designed a semi-active suspension with a PID controller. They optimized PID controller coefficients using an advanced firefly algorithm. They could reduce sprung acceleration amplitude and body acceleration amplitude [21]. Tharehalli Mata et al. proposed a semi-active suspension for a quarter car. They controlled the vibration under random road excitations using a sliding mode controller. They used a particle swarm optimization algorithm to obtain the optimum coefficients for the sliding mode controller [22]. Zhu et al. designed a delay-dependent sliding mode controller for a semi-active suspension system. Their control method is based on a Linear Matrix Inequality [23]. Nguyen et al. created a dynamic model of an active suspension system. They designed and compared different control algorithms for the active suspension system [24]. Li et al. developed an adaptive neural network output-feedback control for an active suspension in case the suspension stiffness is unknown and partial state variables are unmeasurable. They used reinforcement learning to obtain the solution of Hamilton–Jacobi–Bellman equations. They designed the controller with the help of this solution [25]. Nguyen et al. proposed a sliding mode controller for an active suspension. They compared their proposed controller with a PID controller and a passive suspension system. In their simulations using MATLAB, they observed that the sliding mode controller reduced displacement by 14.4% and acceleration by 14.1% compared to the passive suspension system [26]. Viadero-Monasterio et al. developed an $H\infty$ controller for an active suspension under actuator failures. They ensured the stability of the system using the Lyapunov stability approach. They used MATLAB Simulink and Carsim programs for modelling [27].

[1] *Assistant Professor, Electrical and Electronics Engineering Department, TOBB University of Economics and Technology, Ankara, Turkiye. mehmetkarahan@etu.edu.tr

In this study, an active suspension for a quarter car is modeled. The LQR controller is designed for this active suspension. MATLAB/Simulink is used in modeling and simulation. The performance of the designed LQR controller is compared with the PID-controlled active suspension and the classical passive suspension system under different conditions. Comparative analyses were performed under road disturbance, under simultaneous road disturbance and parameter uncertainty and under road disturbance with white noise. Rise time, overshoot, and settling time data of the controllers were analyzed. Simulation results show that the active suspension system with an LQR controller has the least overshoot and fastest settling time under all scenarios among all suspension systems. Thus, simulations prove that the LQR controller has superior performance. The main contributions of this study are as follows:

Most studies in the literature do not perform a stability analysis by generating a pole-zero map of suspension systems. However, in this study, pole-zero maps for the LQR-controlled active suspension, PID-controlled active suspension, and passive suspension were obtained, thus conducting a detailed stability analysis.

Most studies in the existing literature analyze the performance of different suspensions graphically, without comparing numerical data. In this study, time response data (rise time, overshoot, settling time) for the LQR-controlled active suspension, PID-controlled active suspension, and passive suspension under different conditions were obtained and subjected to a comparative robustness analysis.

Most studies in the current literature compare different suspension systems only under instantaneous road disturbances. In this study, suspension systems were examined under road disturbance, under simultaneous road disturbances and parameter uncertainty and under road disturbance with white noise.

In the Materials and Methods section, active and passive suspension models, their mathematical equations and their related parameters are given. In addition, PID controller and LQR controller designs and parameter values of the controllers are explained. In the Results section, stability analysis is performed by performing pole-zero simulations of suspension systems. Then, suspension travel, sprung mass acceleration, and sprung mass motion simulations are performed under road disturbance, under simultaneous road disturbance and parameter uncertainty, and under road disturbance with white noise. Comparative analysis is performed by obtaining time response data. In the Conclusions section, important results of the research are emphasized.

## II. Materials and Methods

In this section, firstly, the modeling of active suspension and passive suspension and their related equations of motion are discussed. Secondly, the equations and controller coefficients for the PID and LQR controllers were explained. The schematic representations of the passive and active suspension are given in Fig. 1. Passive suspension is used for comparison purposes. In Fig. 1 below, $Z_{us}$ represents unsprung mass displacement, $Z_s$ is sprung mass displacement, and $Z_r$ is displacement due to road disturbance.

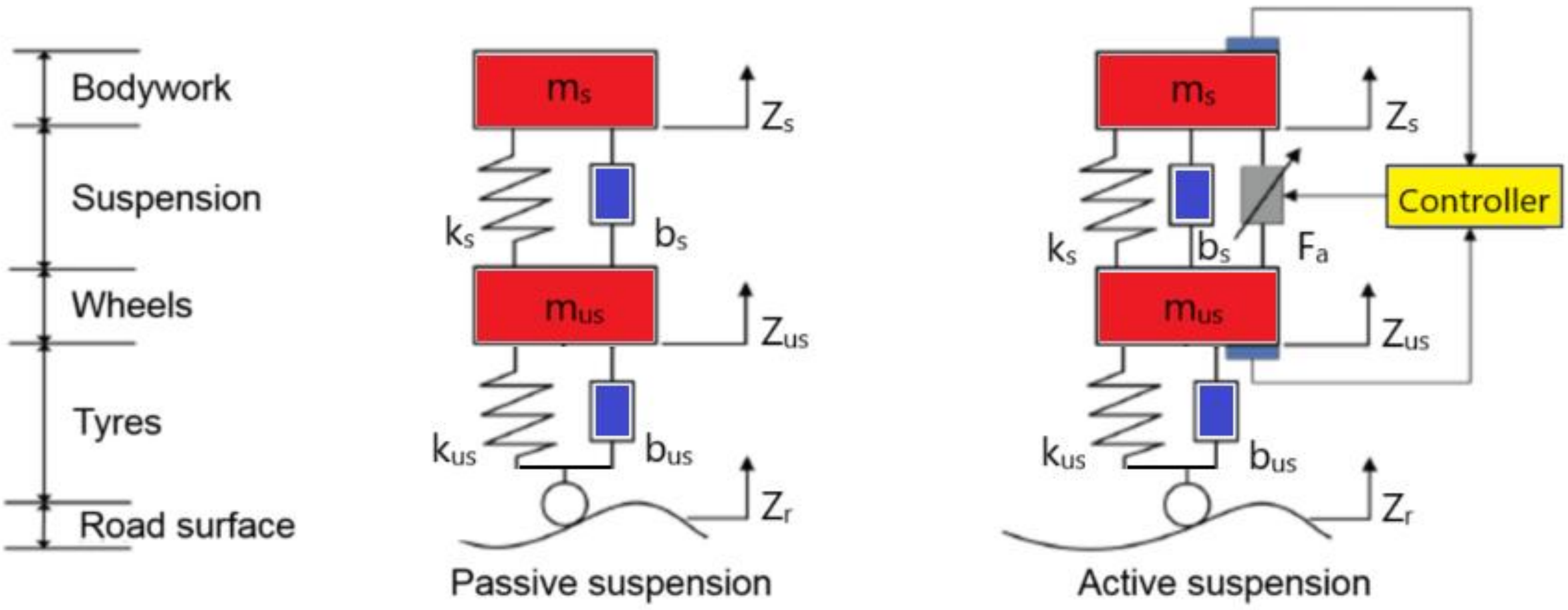

Fig. 1. Schematic representations of the passive and active suspensions

Fig. 2 indicates the difference between sprung mass and unsprung mass. The unsprung mass consists of the vehicle's wheel axles, wheel bearings and hubs, tires, springs, shock absorbers, and suspension links. The sprung mass includes the vehicle's body, chassis, internal components, passengers, and cargo. However, it does not include the components of the unsprung mass [28], [29]. In Fıg. 2, $G_s$ is the center of gravity of the sprung mass, $G_{uf}$ is the center of gravity of the front unsprung mass and $G_{ur}$ is the center of gravity of the rear sprung mass.

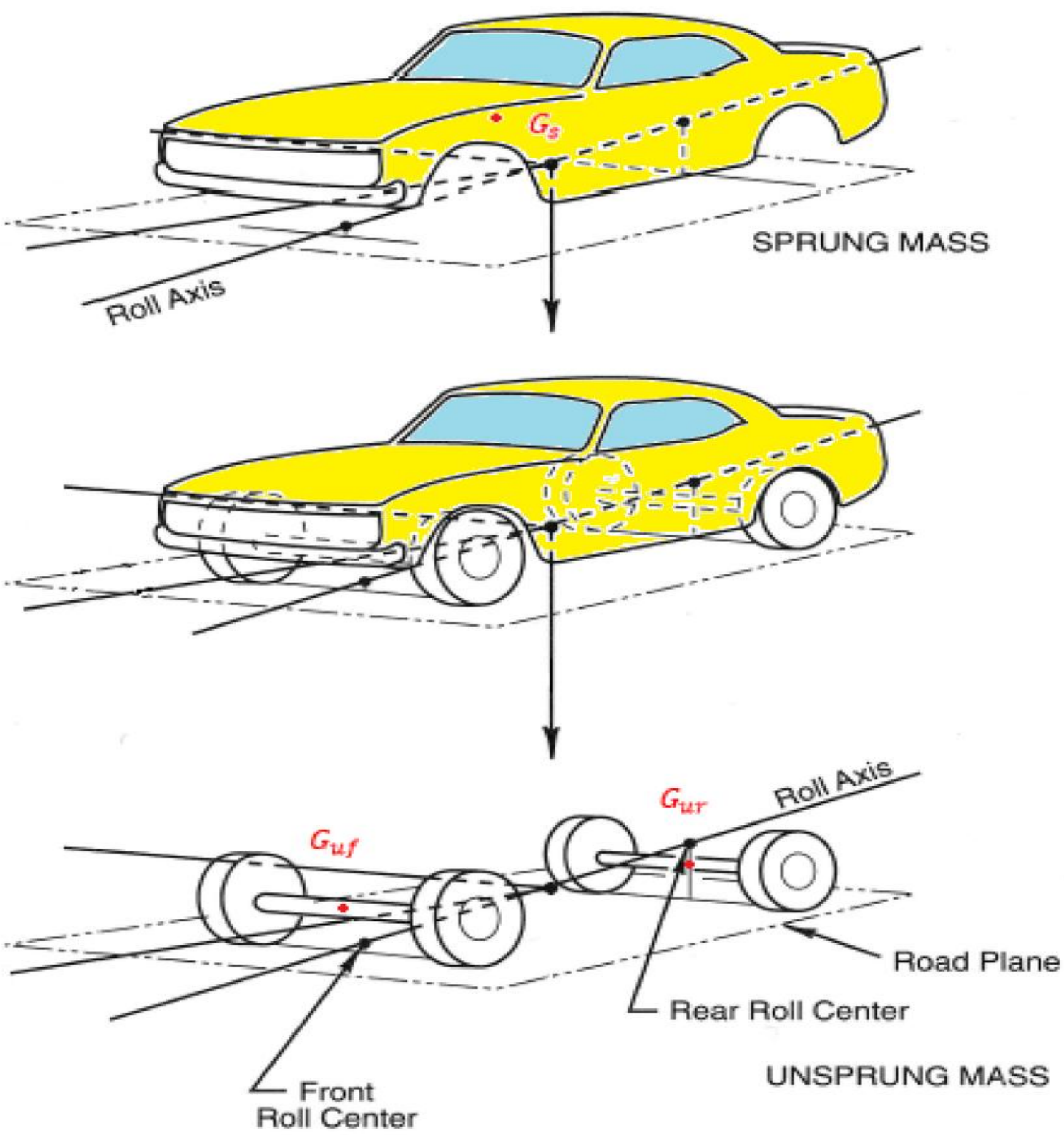

Fig. 2. The difference between sprung mass and unsprung mass

The parameters used in active and passive suspension models are given in Table 1. The wheel damping ratio has been chosen as 0 to provide simplicity.

**Table 1.** Definitions of the suspension parameters

| Parameter | Definition | Value |
|---|---|---|
| $b_s$ | Suspension damping | 1544 Ns/m |
| $b_{us}$ | Tyre damping | 0 Ns/m |
| $k_s$ | Suspension stiffness | 26000 N/m |
| $k_{us}$ | Tyre stiffness | 100000 N/m |
| $m_s$ | Sprung mass | 234 kg |
| $m_{us}$ | Unsprung mass | 43 kg |

The dynamic equations of the quarter car suspension model are as follows:

$$m_s\ddot{Z}_s = b_s\dot{Z}_{us} - b_s\dot{Z}_s - k_s(Z_s - Z_{us}) + F_a \quad (1)$$

$$m_{us}\ddot{Z}_s = b_s\dot{Z}_{us} - b_{us}\dot{Z}_s + b_s\dot{Z}_s + b_{us}\dot{Z}_r - k_s(Z_{us} - Z_s) - k_{us}(Z_{us} - Z_r) - F_a \quad (2)$$

The state space for the active suspension is created using the above equations. The variables of the state space are given in (3). $x_1$ is suspension travel, $x_2$ represents sprung mass velocity, $x_3$ is wheel's deflection, and $x_4$ is wheel's vertical velocity.

$$\left.\begin{array}{c} x_1 = Z_s - Z_{us} \\ x_2 = \dot{Z}_s \\ x_3 = Z_{us} - Z_r \\ x_4 = \dot{Z}_{us} \end{array}\right\} \quad (3)$$

State space representation of the active suspension is given in (4).

$$\begin{bmatrix} \dot{x}_1 \\ \dot{x}_2 \\ \dot{x}_3 \\ \dot{x}_4 \end{bmatrix} = \begin{bmatrix} 0 & 1 & 0 & -1 \\ \frac{-k_s}{m_s} & \frac{-b_s}{m_s} & 0 & \frac{b_s}{m_s} \\ 0 & 0 & 0 & 1 \\ \frac{k_s}{m_{us}} & \frac{b_s}{m_{us}} & \frac{-k_{us}}{m_{us}} & \frac{-(b_s+b_{us})}{m_{us}} \end{bmatrix} \begin{bmatrix} x_1 \\ x_2 \\ x_3 \\ x_4 \end{bmatrix} + \begin{bmatrix} 0 & 0 \\ 0 & \frac{1}{m_s} \\ -1 & 0 \\ \frac{b_{us}}{m_{us}} & \frac{-1}{m_{us}} \end{bmatrix} \begin{bmatrix} \dot{Z}_r \\ F_a \end{bmatrix} \quad (4)$$

After obtaining the state space, the suspension system was modeled using MATLAB Simulink.

### *A. PID Controller*

PID controller and LQR controller designs were made for the active suspension system. In this section, PID controller design is explained. PID control is widely preferred in industry due to its simple structure and easily adjustable coefficients [30], [31]. The block scheme of the PID controller is given in Fig. 3.

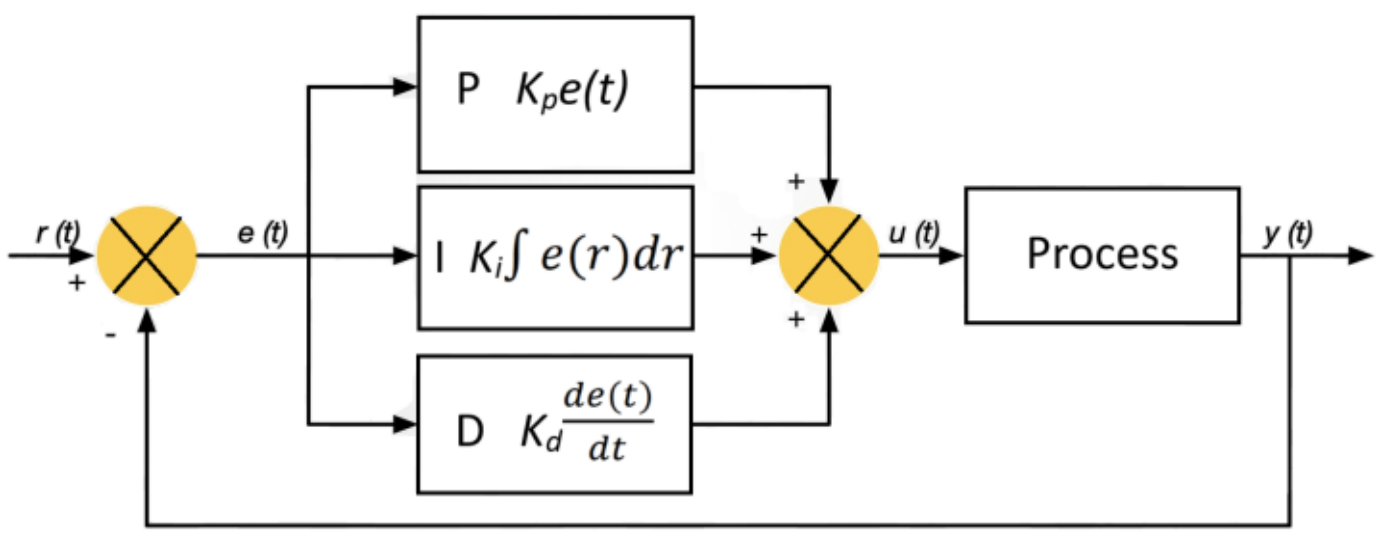

Fig. 3. The block scheme of the PID controller

The equation for the PID controller is given in (5).

$$u(t) = K_p e(t) + K_i \int e(t)dt + K_d \frac{d}{dt} e(t) \quad (5)$$

The e(t) in (5) is the difference in the tracking error. It is found by subtracting the output value y(t) from the reference value r(t) as in (6).

$$e(t) = r(t) - y(t) \quad (6)$$

The values for the PID controller were determined with the PID Tuner feature in Simulink. Two PID controllers were used in the simulations. One PID controller controls the sprung mass motion, and the other controls the suspension travel. The values of the PID controller are given in Table 2.

**Table 2.** PID parameters

| PID Controller Type | $K_p$ | $K_i$ | $K_d$ |
|---|---|---|---|
| Sprung mass motion | $3.2 \times 10^5$ | $5.24 \times 10^3$ | $3.8 \times 10^6$ |
| Suspension travel | 160 | $1.27 \times 10^4$ | 0 |

*B. LQR Controller*

LQR control is an optimum control for dynamic systems at minimum cost [32], [33]. A continuous-time linear system is described as in (7).

$$\dot{x} = Ax + Bu \quad (7)$$

The feedback control law is given in (8). In the equation below, u is the control input and K represents the gain matrix.

$$u = -Kx \quad (8)$$

For an infinite-horizon, continuous-time system, cost function J is defined as in (9) [34].

$$J = \int_0^\infty (x'Qx + u'Ru)dt \quad (9)$$

Q is the state cost matrix, and Q matrix penalizes errors in each state variable. R matrix penalizes the control effort. The gain matrix K is presented in (10).

$$K = R^{-1}B'P \quad (10)$$

The P matrix has to satisfy the reduced matrix equation of Riccati.

$$A'P + PA - PBR^{-1}B'P + Q = 0 \quad (11)$$

The MATLAB Simulink program was used in the design of the LQR controller. Q is a diagonal matrix and R is a positive constant. The Q and R values were calculated by using Bryson's rule in a way that would minimize the quadratic cost function J. Bryson's rule is given in (12-13).

$$Q_{ii} = \frac{1}{maximum\ acceptable\ value\ of\ x_i^2}, i \in \{1,2, \ldots..n\} \quad (12)$$

$$R_{jj} = \frac{1}{maximum\ acceptable\ value\ of\ u_j^2}, j \in \{1,2, \ldots..p\} \quad (13)$$

The Q and R values are defined as in (14-15). By using the lqr function in MATLAB, the gain matrix K is found.

$$Q = \begin{bmatrix} 1x10^{10} & 0 & 0 & 0 \\ 0 & 1x10^8 & 0 & 0 \\ 0 & 0 & 1 & 0 \\ 0 & 0 & 0 & 1 \end{bmatrix} \quad (14)$$

$$R = 1 \tag{15}$$

The general block scheme of the LQR controller is given in Fig. 4.

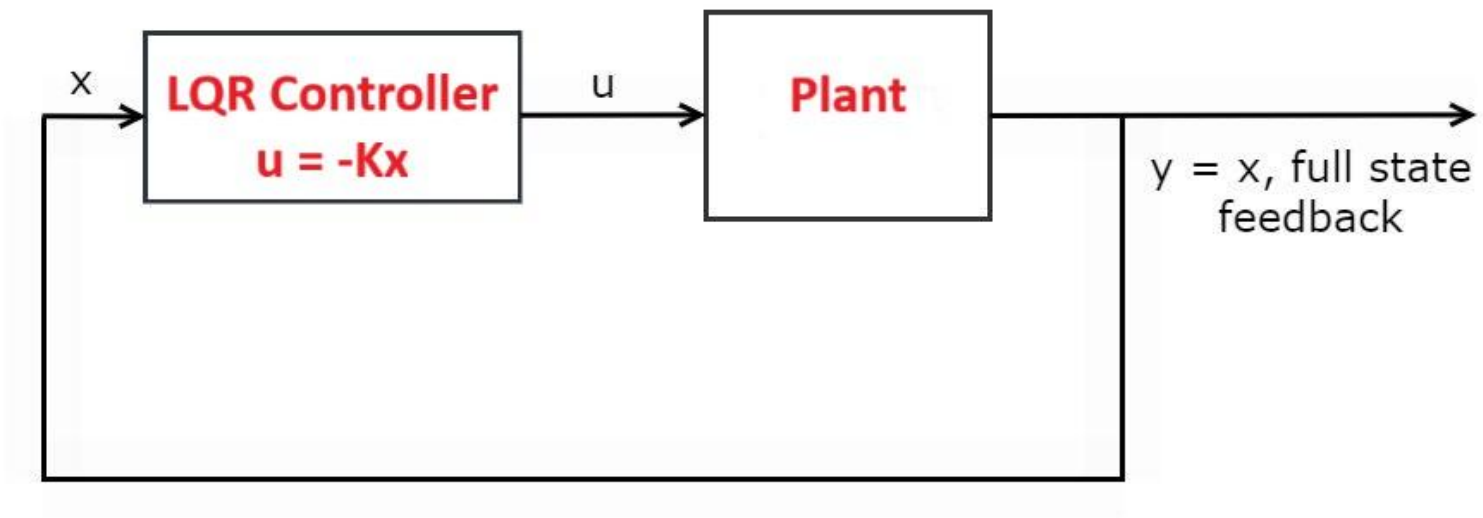

Fig. 4. The general block scheme of the LQR controller

## III. SIMULATIONS

In this section, the results of the simulations performed for the passive suspension system, the PID controller active suspension system, and the LQR controller active suspension system were evaluated. First, the pole-zero maps of the suspension systems were drawn, and stability analysis was performed. Then, suspension travel, sprung mass acceleration, and sprung mass motion simulations were performed under the road disturbance. Active suspension with PID controller, active suspension with LQR controller, and passive suspension were modeled in a single Simulink file, and a comparative analysis was conducted. The Simulink model of the suspension systems is given in Fig. 5.

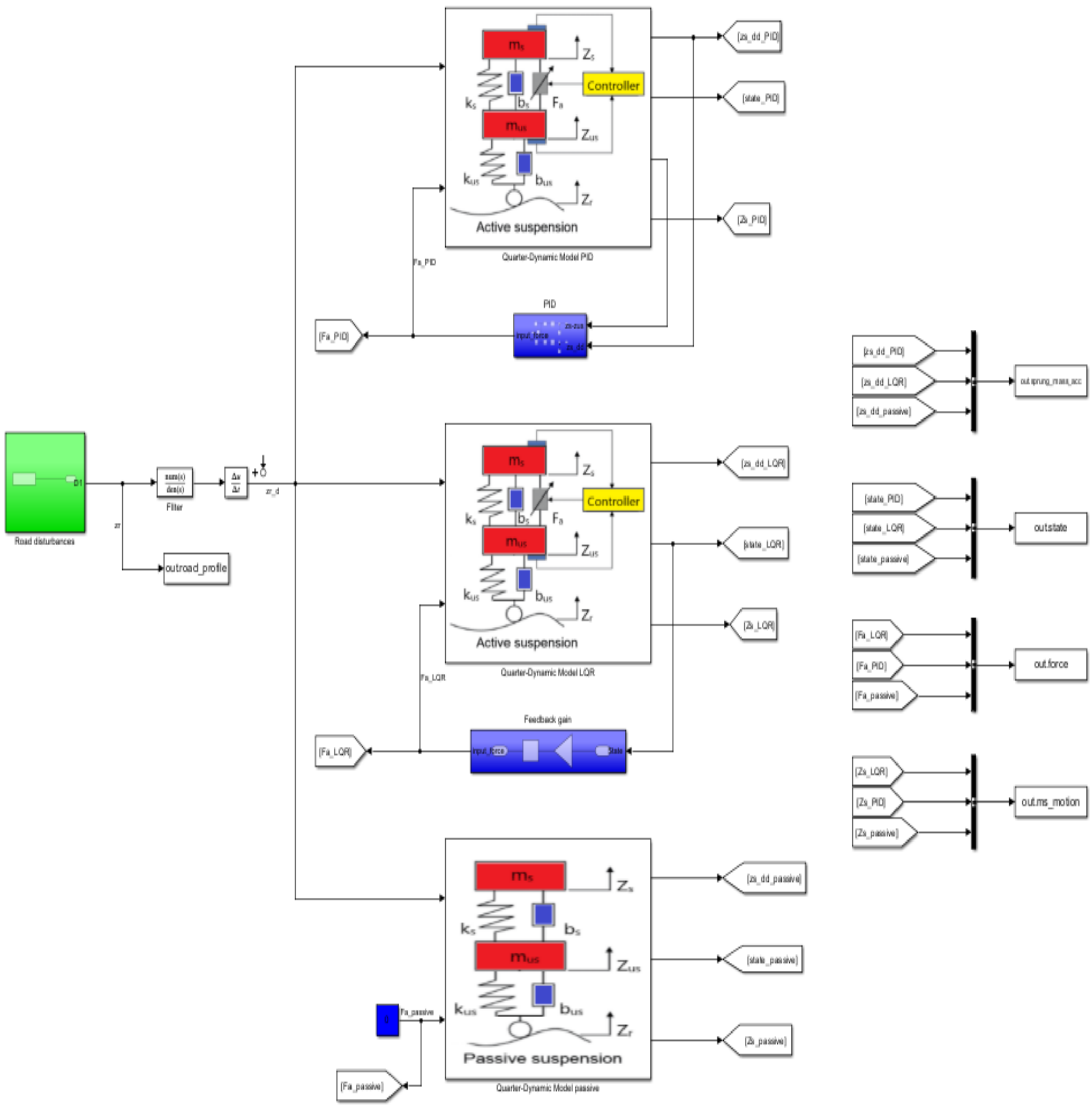

Fig. 5. Simulink models of the suspension systems

### A. *Stability Simulations*

In this section, the zeros and poles of the active suspension with a PID controller, the active suspension with an LQR controller, and the passive suspension were examined. The zero-pole maps of suspension travel, sprung mass acceleration, and sprung mass motion were drawn. In the pole-zero map, the symbols indicated by "x" represent the poles, and the values indicated by "o" represent the zeros. Zeros are the values that make the numerator of the transfer function of a system 0. Poles are the values that make the denominator of the transfer function of a system 0. In order for a system to be stable, all pole values must be to the left of the imaginary axis. Stability analysis was performed by checking whether the poles were to the left of the imaginary axis. Fig. 6 represents the pole-zero map of the suspension travel. The poles of the passive suspension system, the active suspension system with PID controller, and the active suspension system with LQR controller are located to the left of the imaginary axis. In this case, all three suspension systems are stable in the suspension travel pole zero map.

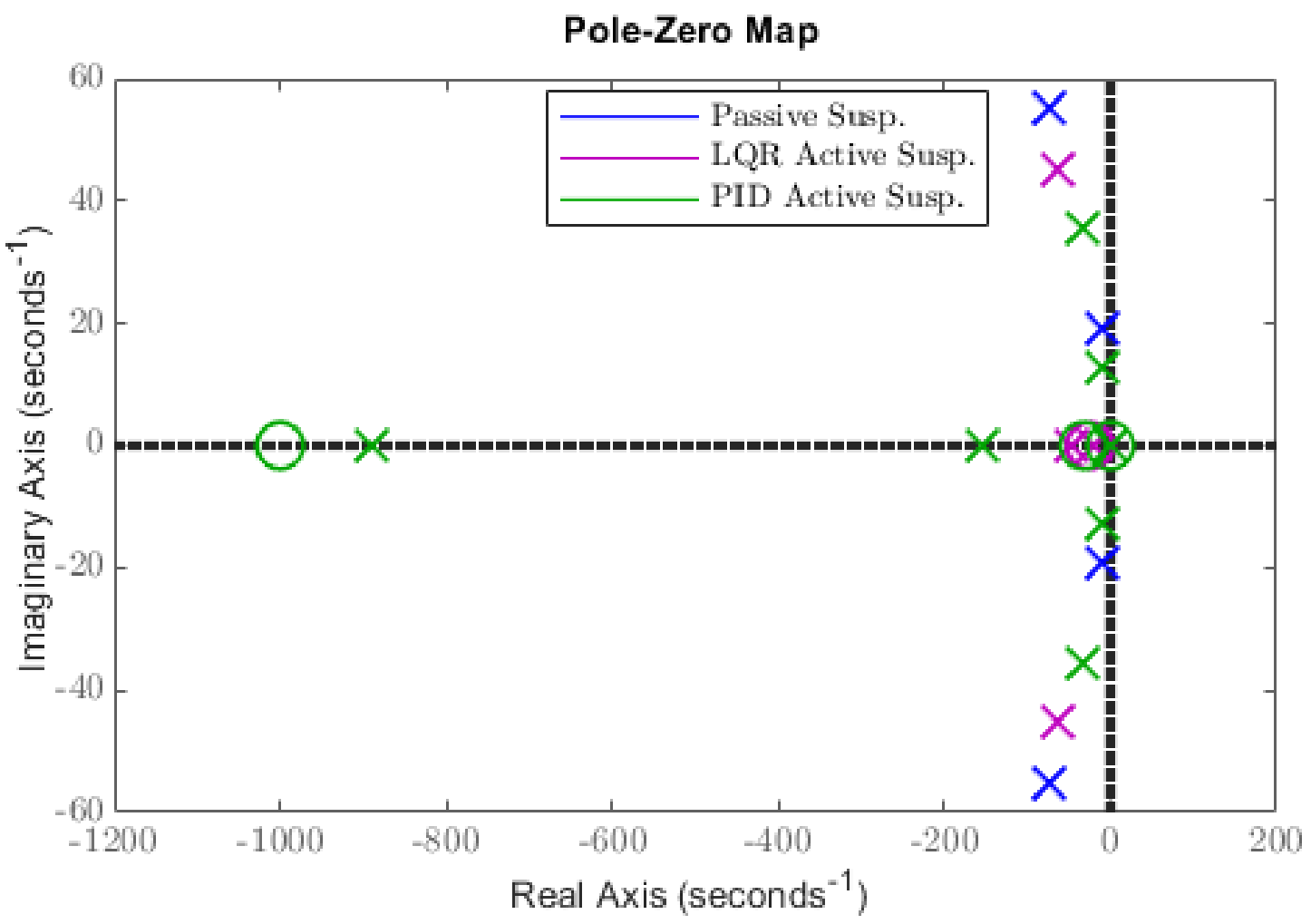

Fig. 6. Poles and zeros of the suspension travel

Fig. 7 shows the pole-zero map of the sprung mass acceleration. All poles of the passive suspension system, the active suspension with PID controller, and the active suspension with LQR controller are to the left of the imaginary axis. In this case, all three suspension systems are stable.

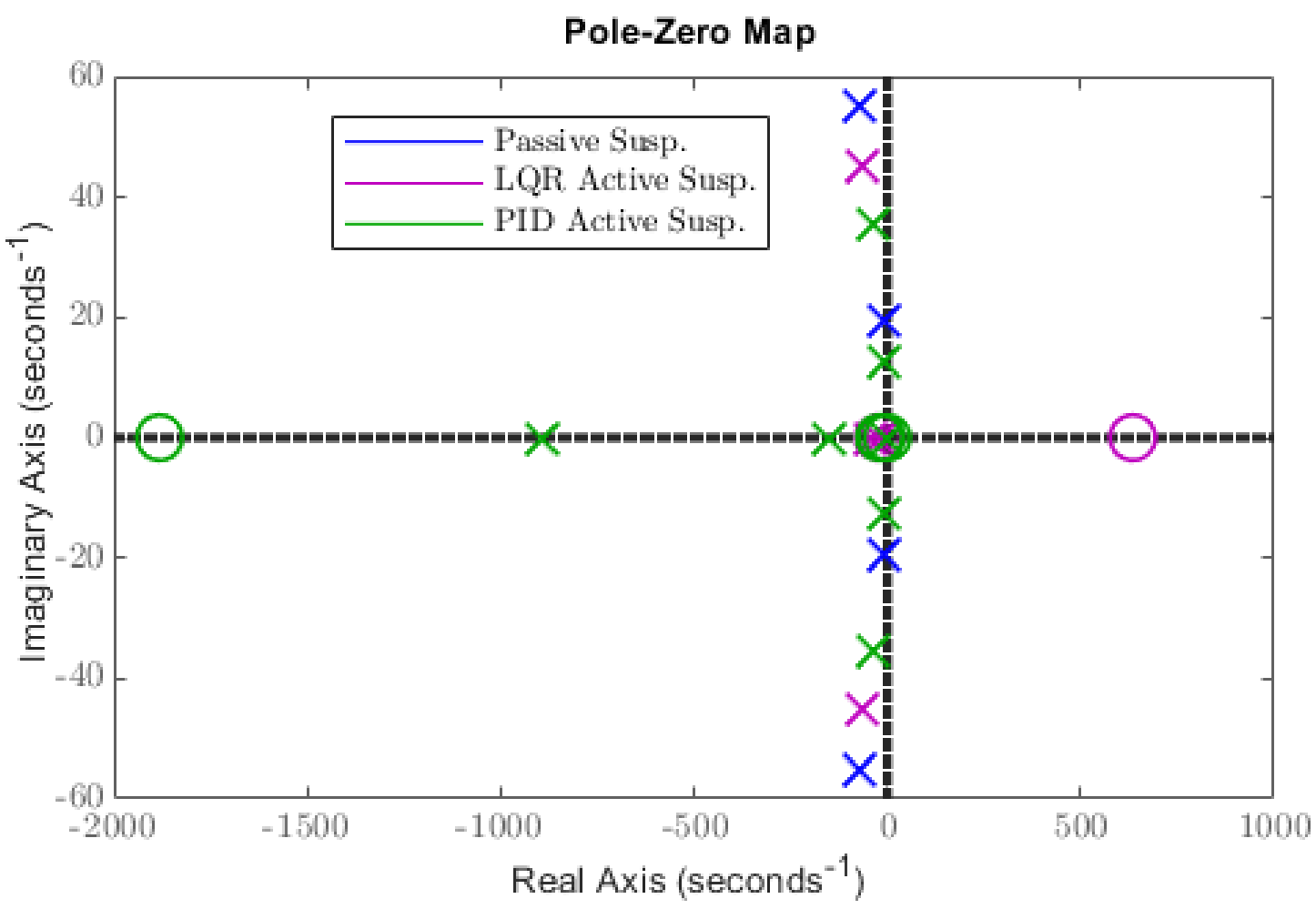

Fig. 7. Poles and zeros of the sprung mass acceleration

Fig. 8 presents the pole-zero map of the sprung mass motion. When the Figure 8 is examined, it is seen that all poles of the suspension systems are located to the left of the imaginary axis. In this case, all three suspension systems are stable.

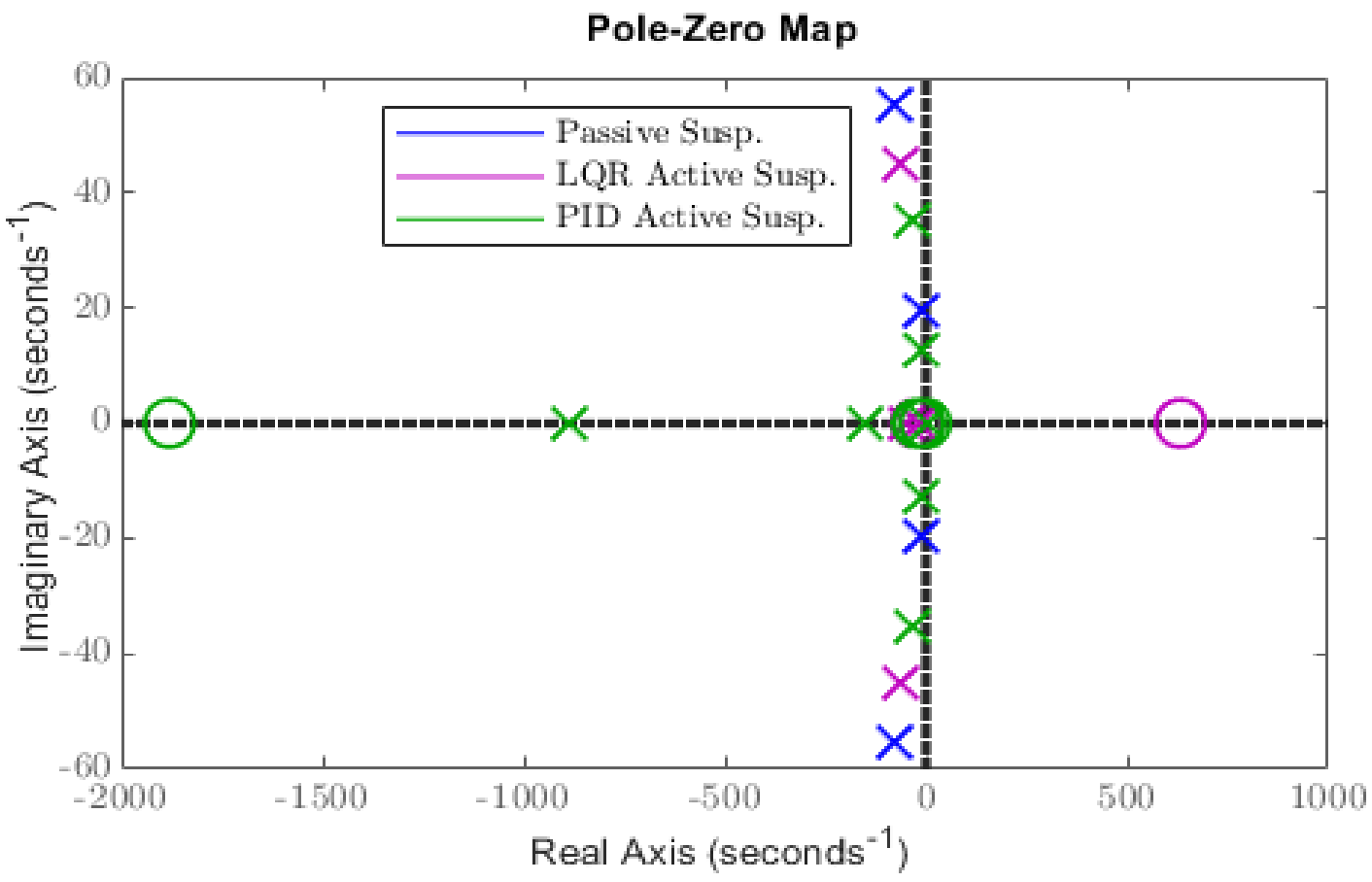

Fig. 8. Poles and zeros of the sprung mass motion

### B. *Suspension Simulations under Road Disturbance*

In the simulations, the vehicle speed was assumed to be 72 km/h. A disturbance was given to the road in the 1st second of the simulation. The amplitude of the applied disturbance is 0.08 m. The simulation of this disturbance is shown in Fig. 9.

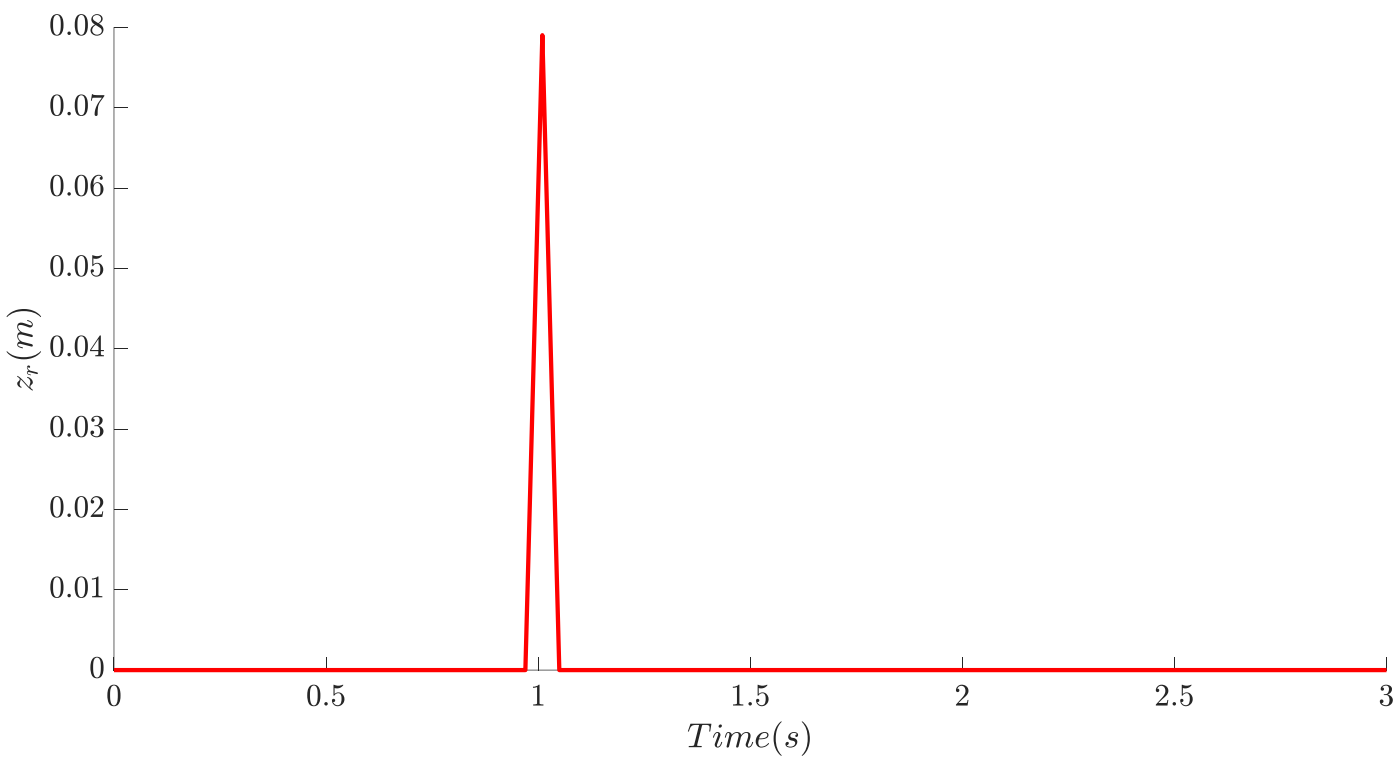

Fig. 9. Road disturbance

After the disturbance is applied to the road, suspension travel, sprung mass acceleration, and sprung mass motion simulations are performed. In the simulations performed, time response data of passive suspension, PID-controlled active suspension, and LQR-controlled active suspension are obtained. Then, the time response data (rise time, overshoot, and settling time) of the suspension systems were compared. The rise time is taken as the time to reach 90% of the reference value. The rise time is in seconds. The overshoot shows how much the reference value was exceeded. The over-shoot value was taken as m for suspension travel simulation, m/s2 for sprung mass acceleration simulation, and m for suspension motion. The settling time is taken as the time to settle within 2% of the reference value. The settling time is in seconds. Fig. 10 shows the suspension travel simulation of the passive suspension, the PID-controlled active suspension, and the LQR-controlled active suspension.

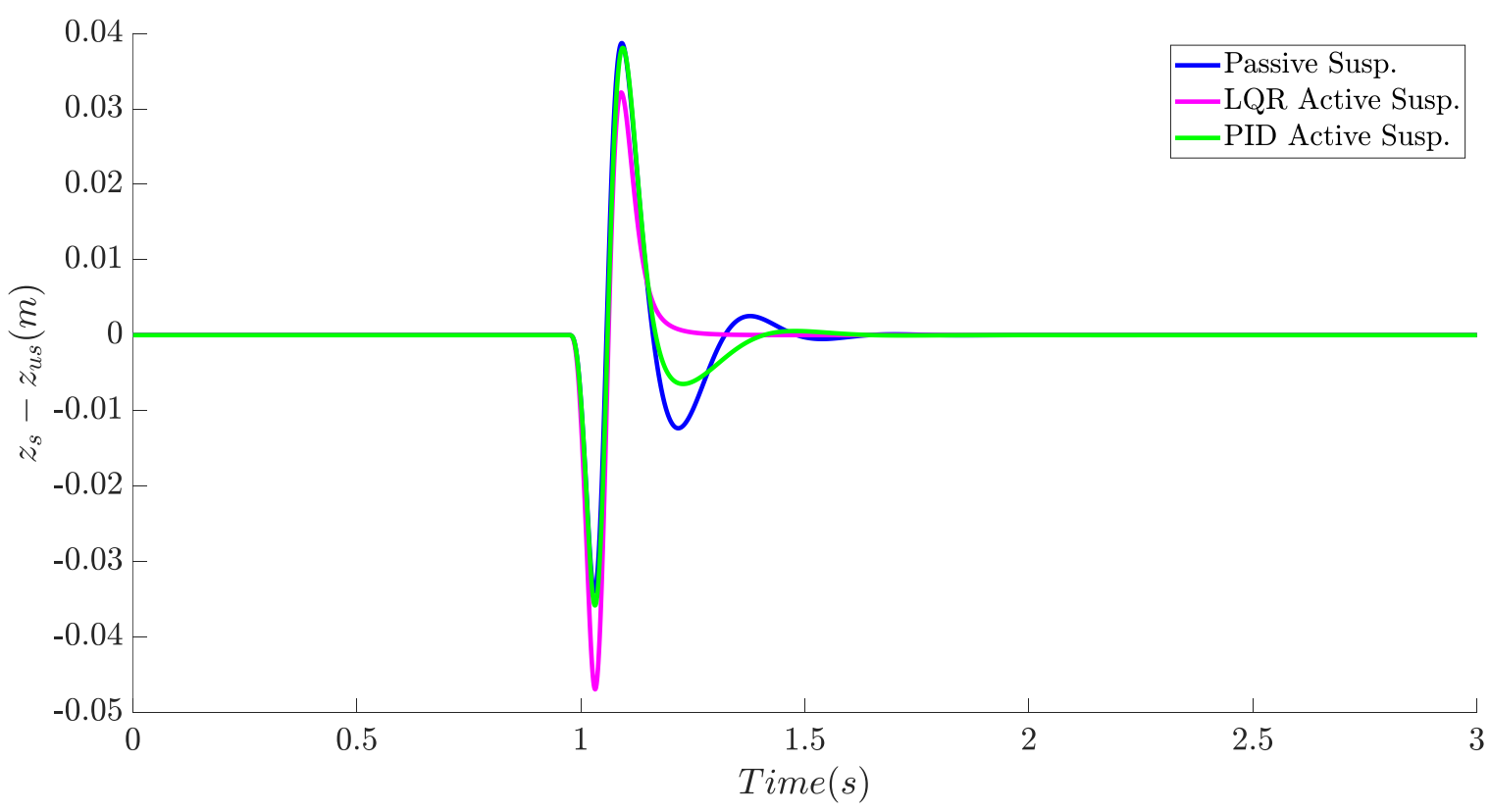

Fig. 10. Suspension travel simulation under road disturbance

Table 3 shows the time response data of passive suspension, PID-controlled active suspension, and LQR-controlled active suspension.

**Table 3.** Time response for suspension travel under road disturbance

| Suspension type | Rise time (s) | Overshoot (m) | Settling time (s) |
|---|---|---|---|
| Passive | 1.29 s | 0.0387 m | 1.451 s |
| PID-controlled active | 1.30 s | 0.038 m | 1.377 s |
| LQR-controlled active | 1.15 s | -0.047 m | 1.21 s |

The rise time of passive suspension and the rise time of an active suspension with a PID controller are very close to each other. Passive suspension has a rise time of 0.01 seconds shorter than active suspension with a PID controller. Active suspension with an LQR controller has the fastest rise time. It has a rise time of approximately 0.15 seconds, shorter than passive suspension and active suspension with a PID controller. Active suspension with an LQR controller has the highest negative overshoot. However, it oscillates less and settles to the reference value quickly. Passive suspension has the highest positive overshoot. Active suspension with a PID controller has less oscillation than passive suspension. The passive suspension oscillation lasts much longer than LQR and PID-controlled active suspensions. When a disturbance is given to the road in the 1st second, the passive suspension starts to oscillate and settles within 2% of the reference in the 1.451st second. The PID-controlled active suspension settles within 2% of the reference in 1.377th seconds. The LQR-controlled active suspension settles within 2% of the reference in 1.21 seconds. Thus, it is revealed that the LQR-controlled suspension oscillates less than other suspensions and offers a more comfortable ride.

The sprung mass acceleration simulation of different suspension systems under road disturbance is shown in Fig. 11.

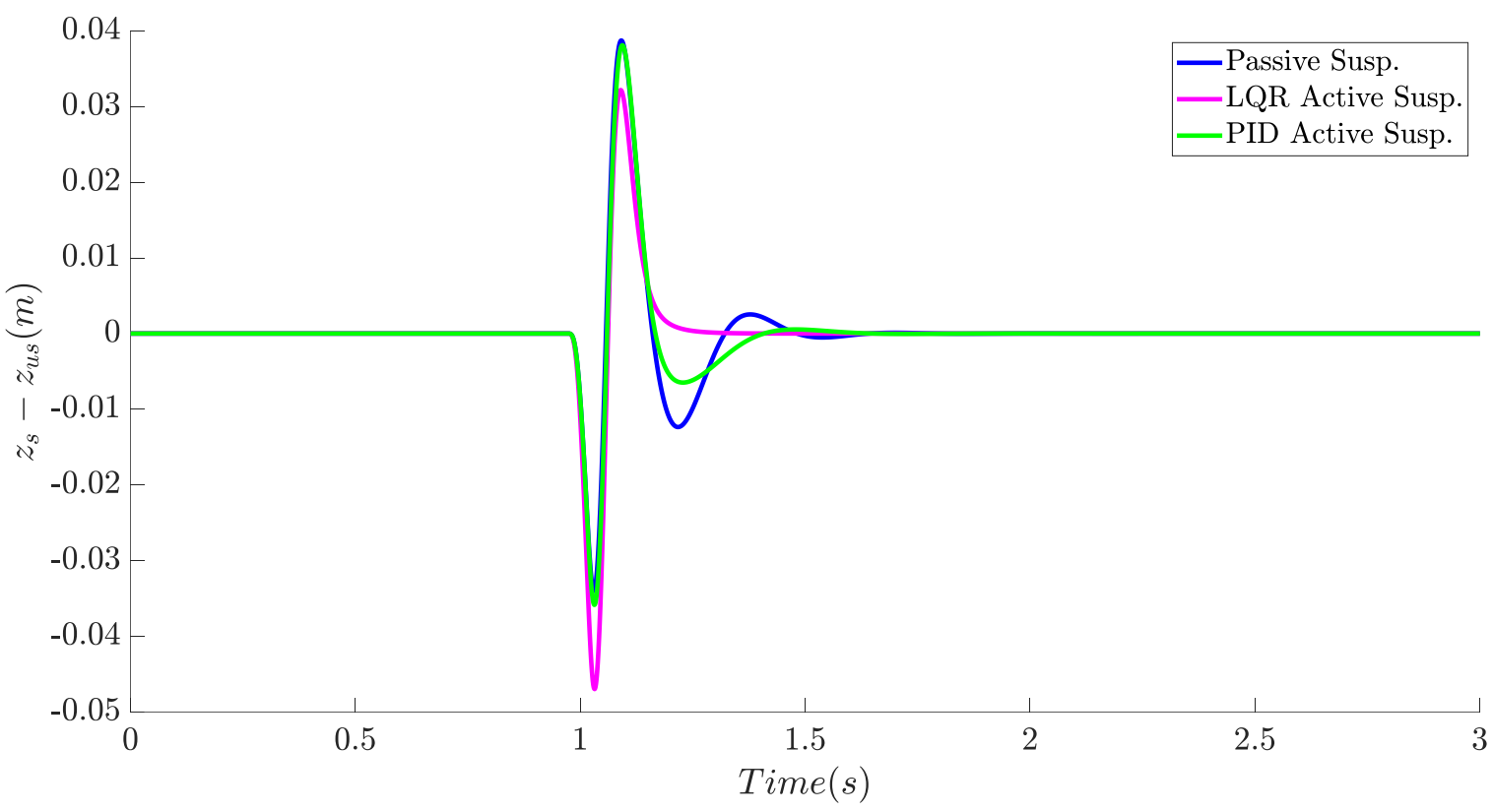

Fig. 11. Sprung mass acceleration simulation under road disturbance

Table 4 shows the time response of suspensions for sprung mass acceleration. Active suspension with an LQR controller has the fastest rise time of 1.151 s. Active suspension with a PID controller has the second-fastest rise time of 1.174 s. Passive suspension has the longest rise time of 1.216 s. The LQR controller has the least overshoot. The overshoot values of passive suspension and active suspension with a PID controller are very close to each other. However, active suspension with a PID controller shows a slightly higher overshoot value. Active suspension with an LQR controller shows the fastest settling time. Active suspension with a PID controller has a faster settling time than passive suspension. Passive suspension has the longest settling time. Simulation results show that active suspension with an LQR controller is the most successful suspension in all respects.

**Table 4.** Time response for sprung mass acceleration under road disturbance

| Suspension type | Rise time (s) | Overshoot (m/s$^2$) | Settling time (s) |
|---|---|---|---|
| Passive | 1.216 s | -55.1 m/s$^2$ | 1.378 s |
| PID-controlled active | 1.174 s | -56.1 m/s$^2$ | 1.238 s |
| LQR-controlled active | 1.151 s | -47.54 m/s$^2$ | 1.191 s |

Sprung mass motion simulation of three different suspension systems under road disturbance is presented in Fig. 12.

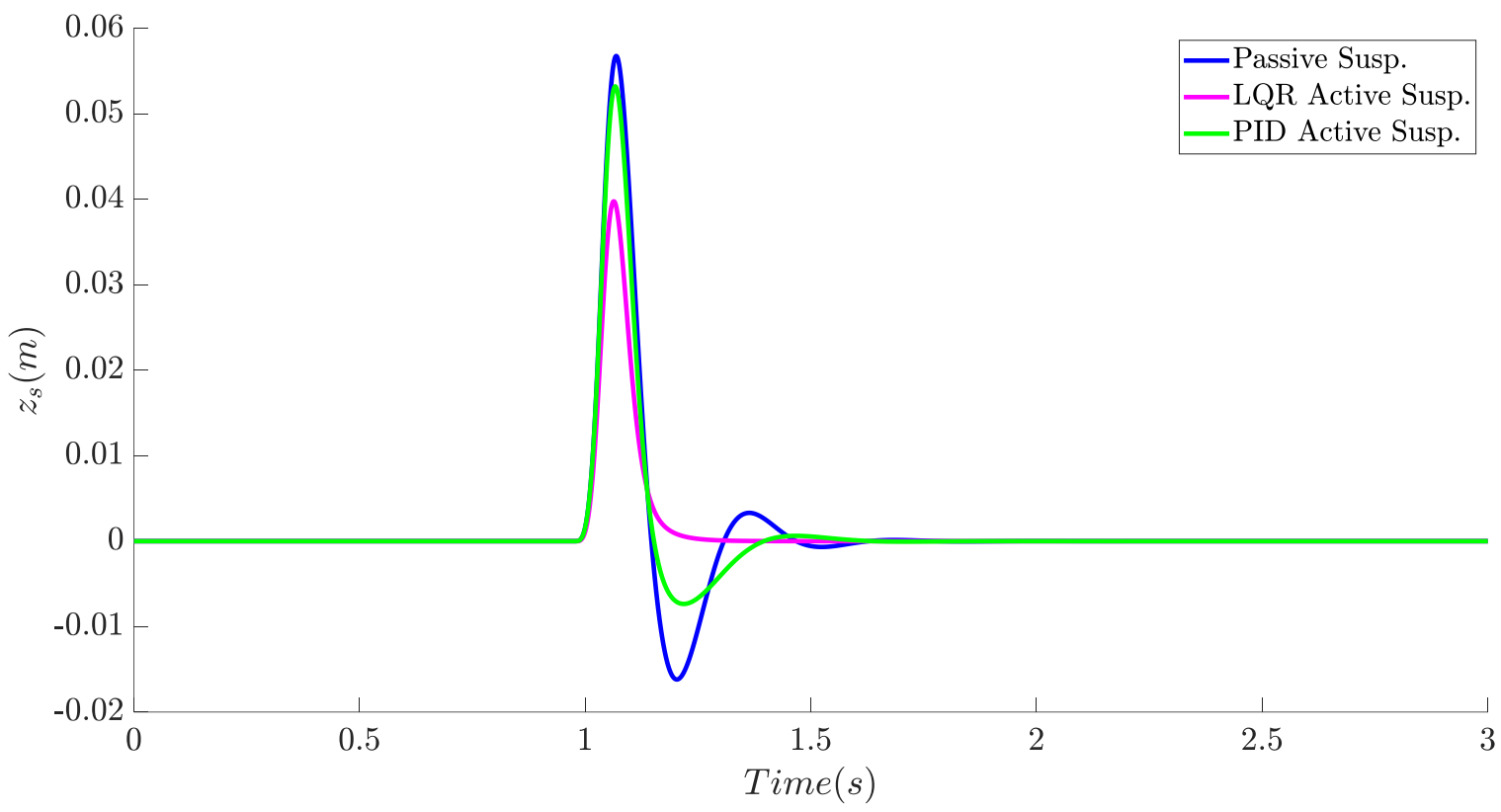

Fig. 12. Sprung mass motion simulation under road disturbance

Table 5 presents the time response of suspensions for sprung mass motion. When the simulation data in Table 5 is examined, it is observed that the passive suspension and the active suspension with a PID controller have almost the same rise time of 1.138 s and 1.139 s. The LQR controller has a longer rise time of 1.15 s. The passive suspension has the highest overshoot with 0.0567 m. The active suspension with a PID controller shows the second-highest overshoot with 0.0532 m. The active suspension with the LQR controller is the suspension system with the least overshoot, with 0.039 m. The passive suspension has the longest settling time with 1.428 s. The active suspension with a PID controller has the second-longest settling time with 1.346 s. The active suspension system with LQR controller shows the shortest settling time with 1.204 s. When all these results are evaluated together, it is clear that the active suspension system with the LQR controller is the suspension with the least overshoot and the fastest settling time.

**Table 5.** Time response for sprung mass motion under road disturbance

| Suspension type | Rise time (s) | Overshoot (m) | Settling time (s) |
|---|---|---|---|
| Passive | 1.138 s | 0.0567 m | 1.428 s |
| PID-controlled active | 1.139 s | 0.0532 m | 1.346 s |
| LQR-controlled active | 1.15 s | 0.039 m | 1.204 s |

### *C. Suspension Simulations under Road Disturbance and Parameter Uncertainty*

The weight of the sprung mass may increase depending on the number of passengers and the amount of load inside the vehicle. The increase in the ratio of the sprung mass to the unsprung mass is a factor that makes the control of the vehicle more difficult. In this section, considering that the weight of the sprung mass may increase, a parameter uncertainty of +20% was applied to the sprung mass, and simulations were performed. The weight of the sprung mass for the quarter car was taken as 281 kg with a 20% increase. Fig. 13 shows the suspension travel simulation under road disturbance and parameter uncertainty.

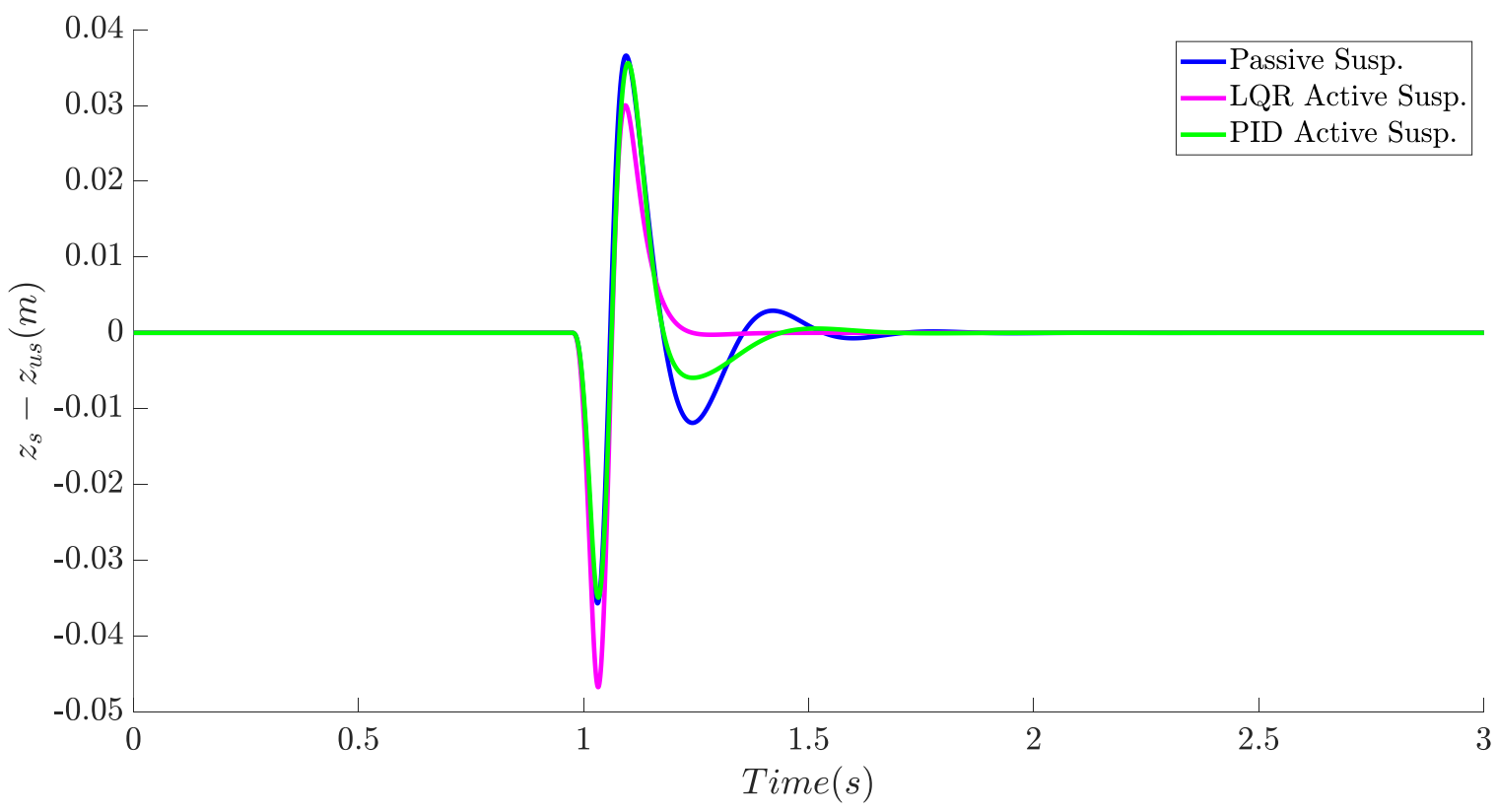

Fig. 13. Suspension travel under road disturbance and parameter uncertainty

The time response of the suspension travel under parameter uncertainty is given in Table 6. Active suspension with an LQR controller has the fastest rise time. Passive suspension and active suspension with a PID controller have very similar rise times, but passive suspension has a slightly faster rise time. Active suspension with an LQR controller has the highest negative overshoot but the lowest positive overshoot. Passive suspension has the highest positive overshoot. Active suspension with a PID controller has the second-highest positive overshoot. Passive suspension has the longest settling time. Active suspension with a PID controller has the second-highest settling time. Active suspension with an LQR controller has the shortest settling time.

**Table 6.** Time response for suspension travel under road disturbance and parameter uncertainty

| Suspension type | Rise time (s) | Overshoot (m) | Settling time (s) |
|---|---|---|---|
| Passive | 1.323 s | 0.0365 m | 1.507 s |
| PID-controlled active | 1.329 s | 0.0356 m | 1.406 s |
| LQR-controlled active | 1.172 s | -0.0467 m | 1.211 s |

Fig. 14 shows the sprung mass acceleration simulation under road disturbance and parameter uncertainty.

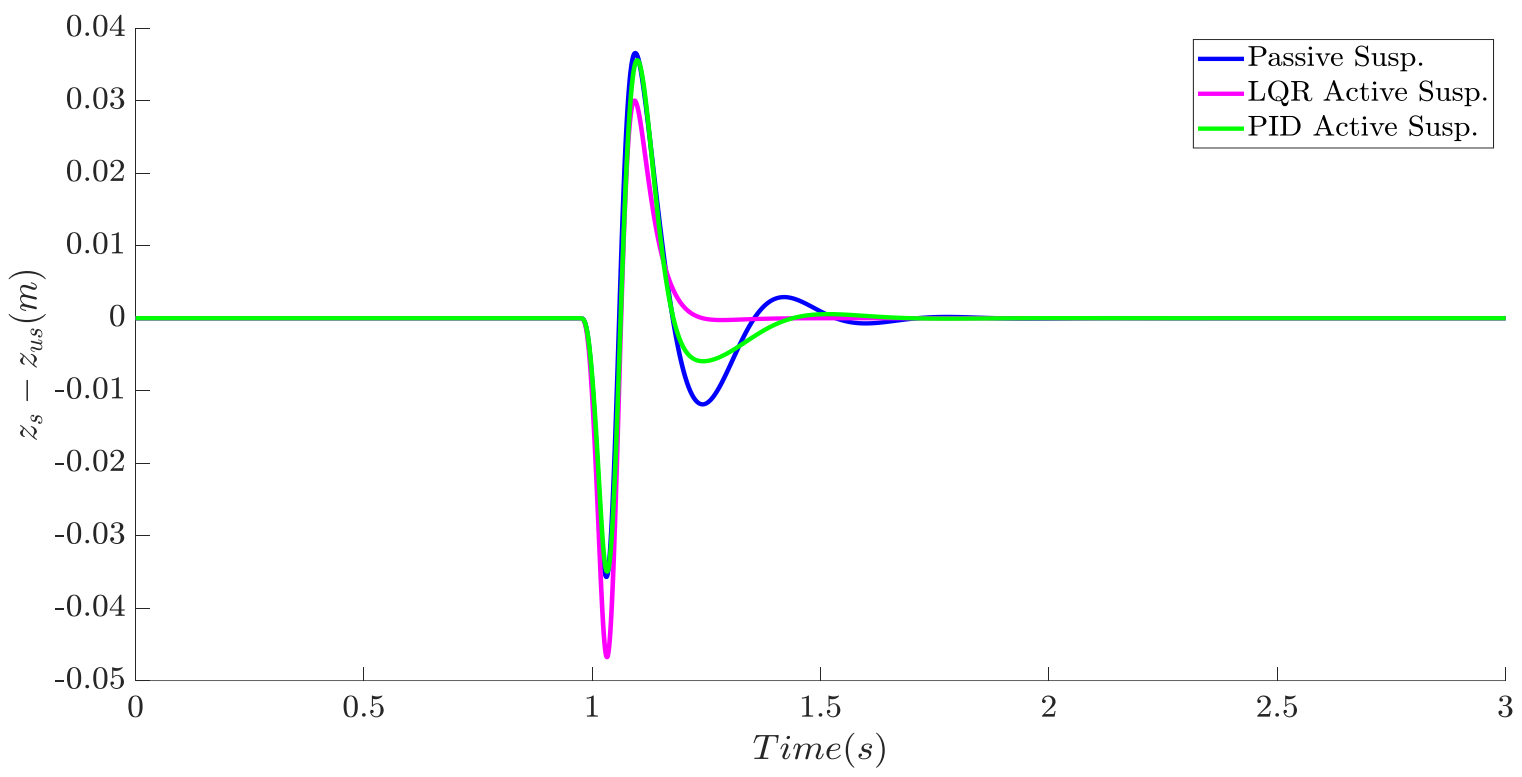

Fig. 14. Sprung mass acceleration under road disturbance and parameter uncertainty

The time response of the sprung mass acceleration under parameter uncertainty is represented in Table 7. Active suspension with an LQR controller has the fastest rise time. Active suspension with a PID controller has the second-fastest rise time. Passive suspension has the longest rise time. Active suspension with a PID controller shows the highest overshoot value. Passive suspension shows the second-highest overshoot value. Active suspension with an LQR controller shows the least overshoot. Passive suspension has the longest settling time. Active suspension with a PID controller has the second-longest settling time. Active suspension with an LQR controller has the shortest settling time. When all the results are evaluated together, it is clear that the active suspension with the LQR controller is the suspension that rises the fastest, shows the least overshoot, and settles the fastest.

**Table 7.** Time response for sprung mass acceleration under road disturbance and parameter uncertainty

| Suspension type | Rise time (s) | Overshoot (m/s$^2$) | Settling time (s) |
|---|---|---|---|
| Passive | 1.241 s | -43.83 m/s$^2$ | 1.428 s |
| PID-controlled active | 1.181 s | -48.61 m/s$^2$ | 1.245 s |
| LQR-controlled active | 1.161 s | -39.96 m/s$^2$ | 1.222 s |

Fig. 15 shows the sprung mass motion simulation of the suspension systems under simultaneous road disturbance and parameter uncertainty.

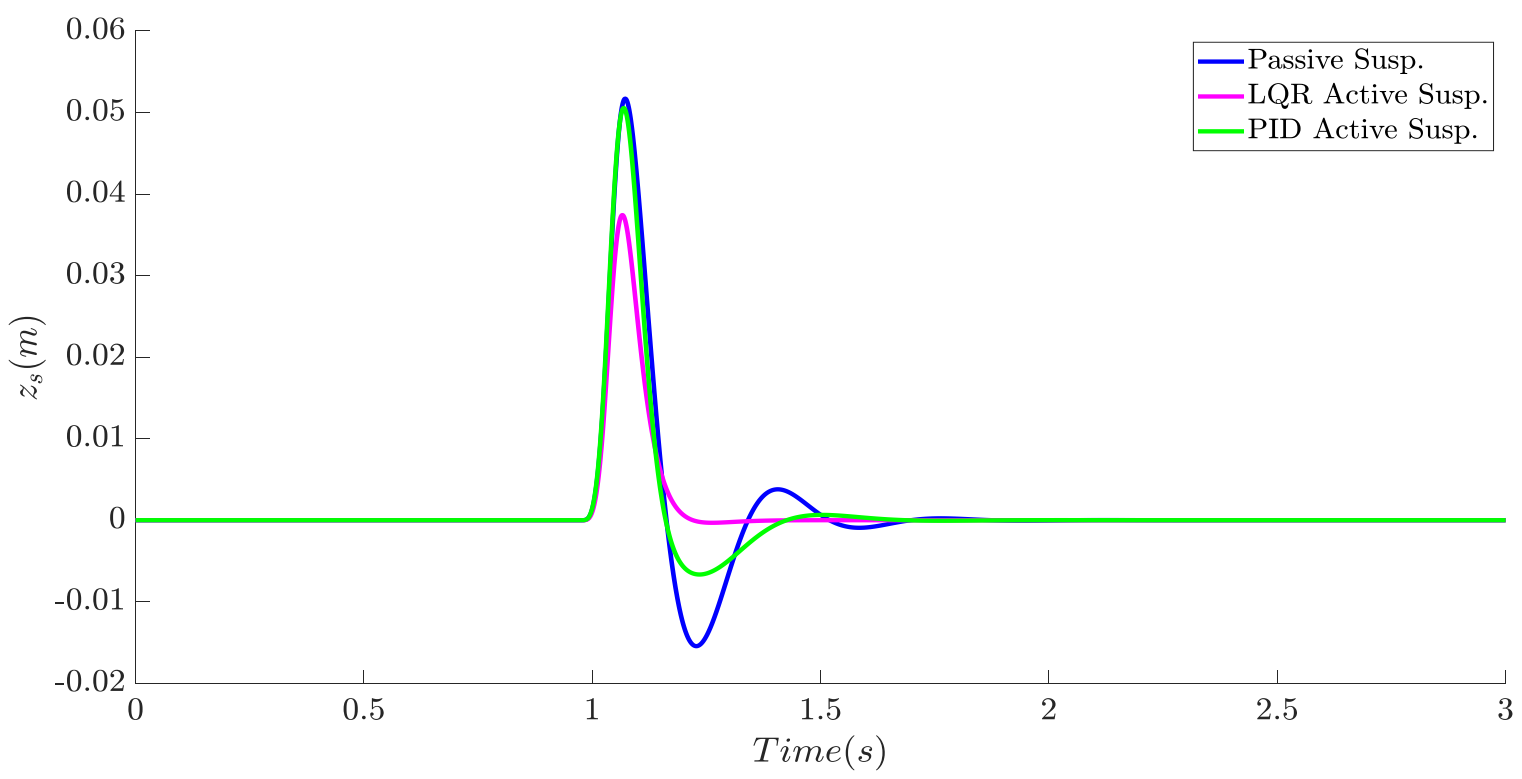

Fig. 15. Sprung mass motion under road disturbance and parameter uncertainty

The time response of the sprung mass motion under parameter uncertainty is represented in Table 8. Passive suspension has the longest rise time. Active suspension with a PID controller has the second-longest rise time. Active suspension with an LQR controller has the fastest rise time. Passive suspension shows the highest overshoot. Active suspension with a PID controller has the second-highest overshoot. Active suspension with an LQR controller shows the least overshoot. Passive suspension has the longest settling time. Active suspension with a PID controller has the second-longest settling time. Active suspension with an LQR controller has the shortest settling time. In all-time response data, the LQR controller suspension shows a clear superiority over the others.

**Table 8.** Time response for sprung mass motion under road disturbance and parameter uncertainty

| Suspension type | Rise time (s) | Overshoot (m) | Settling time (s) |
|---|---|---|---|
| Passive | 1.307 s | 0.0517 m | 1.491 s |
| PID-controlled active | 1.297 s | 0.050 m | 1.387 s |
| LQR-controlled active | 1.206 s | 0.037 m | 1.198 s |

*D. Suspension Simulations under Road Disturbance and Band-Limited White Noise*

In this section, simulations were performed with the addition of band-limited white noise for road disturbances. This allows the performance of the suspensions to be evaluated against road disturbances that persist throughout the simulation, rather than against sudden disturbances that appear for the first second. The noise power of the band-limited white Gaussian noise was $1\times10^{-5}$, and the sampling time was 0.1. Because the noise applied in the simulation increases oscillations, the settling time was taken to be within ±5% of the reference value. The road disturbance after the noise is applied is shown in Fig. 16.

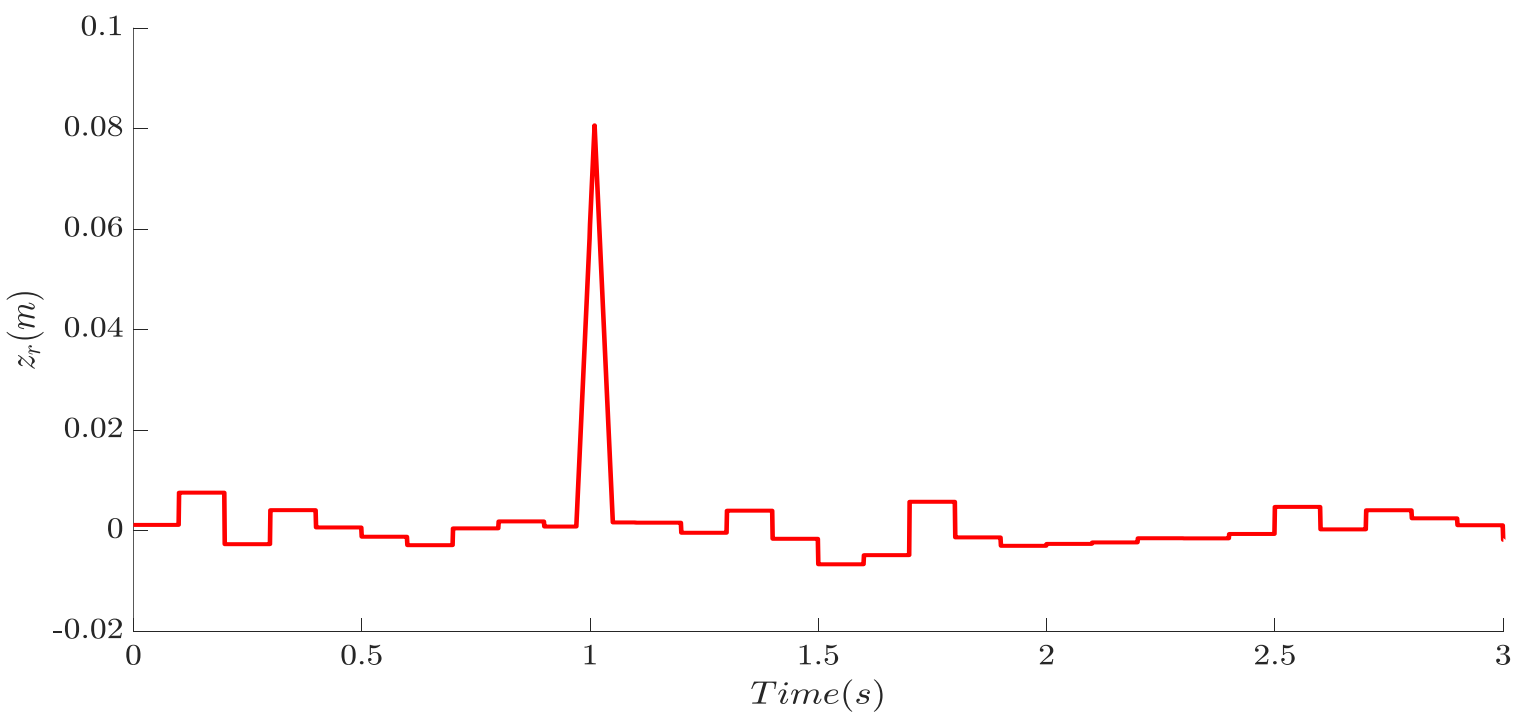

Fig. 16. Road disturbance with band-limited white noise

Fig. 17 shows the suspension travel simulation under the road disturbance after the band-limited white noise is applied.

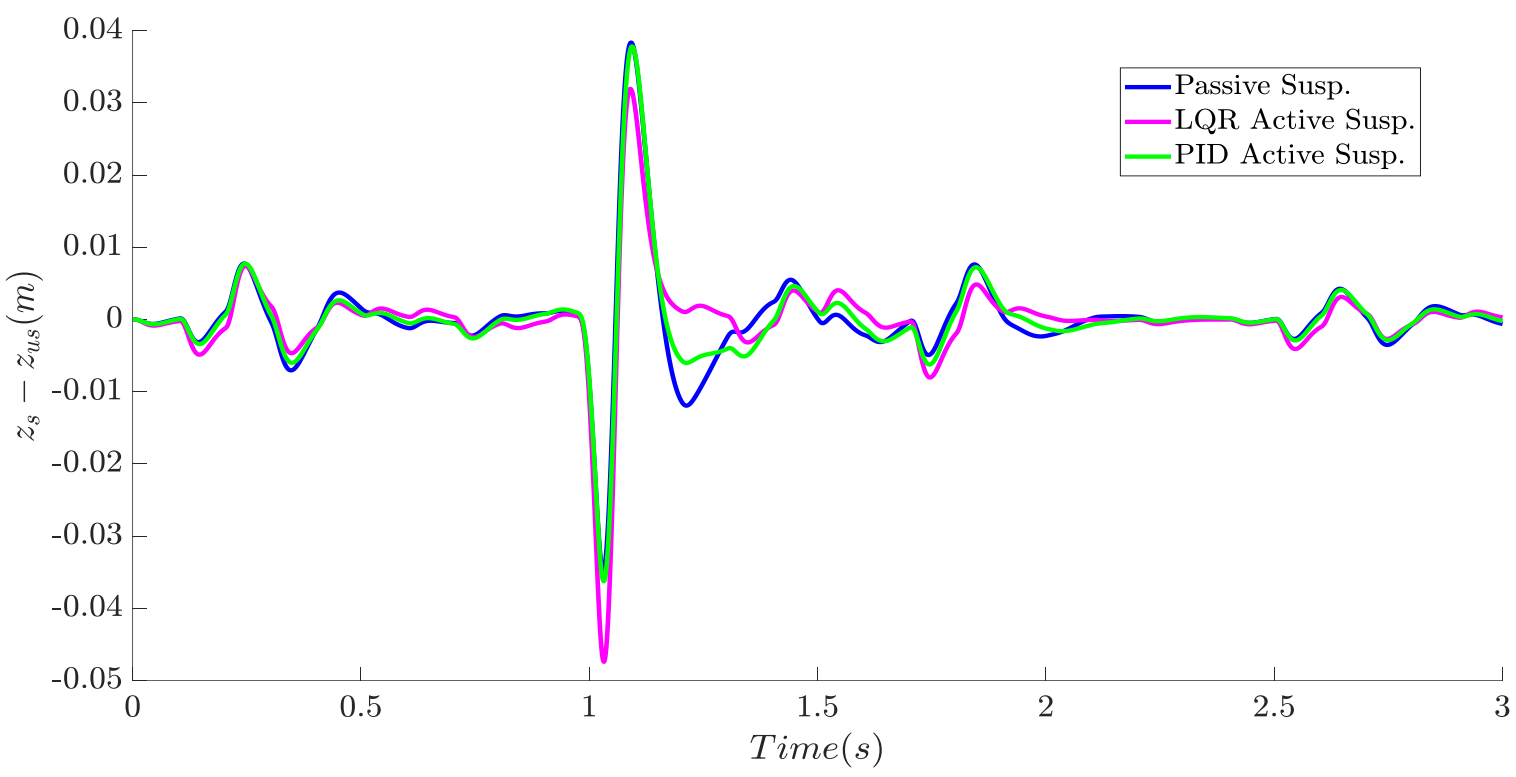

Fig. 17. Suspension travel under road disturbance with band-limited white noise

Time response data for suspension travel simulation under road disturbance with band-limited white noise are presented in Table 9.

**Table 9.** Time response for suspension travel under road disturbance and parameter uncertainty

| Suspension type | Rise time (s) | Overshoot (m) | Settling time (s) |
|---|---|---|---|
| Passive | 1.291 s | 0.038 m | 2.783 s |
| PID-controlled active | 1.363 s | 0.037 m | 2.777 s |
| LQR-controlled active | 1.159 s | -0.047 m | 2.578 s |

Examining Table 9, it is observed that the active suspension with the LQR controller has the fastest rise time. Passive suspension comes second. Active suspension with a PID controller has the longest rise time. Passive suspension has the highest positive overshoot. Active suspension with a PID controller has the highest positive overshoot. Active suspension with an LQR controller has the highest negative overshoot. Passive suspension has the longest settling time. Active suspension with a PID controller has the second-longest settling time. Active suspension with an LQR controller has the shortest settling time. The LQR-controlled active suspension clearly provides a more comfortable ride because it has the fastest rise time and the shortest settling time.

Sprung mass acceleration simulation of the suspension systems under road disturbance with white noise is given in Fig. 18.

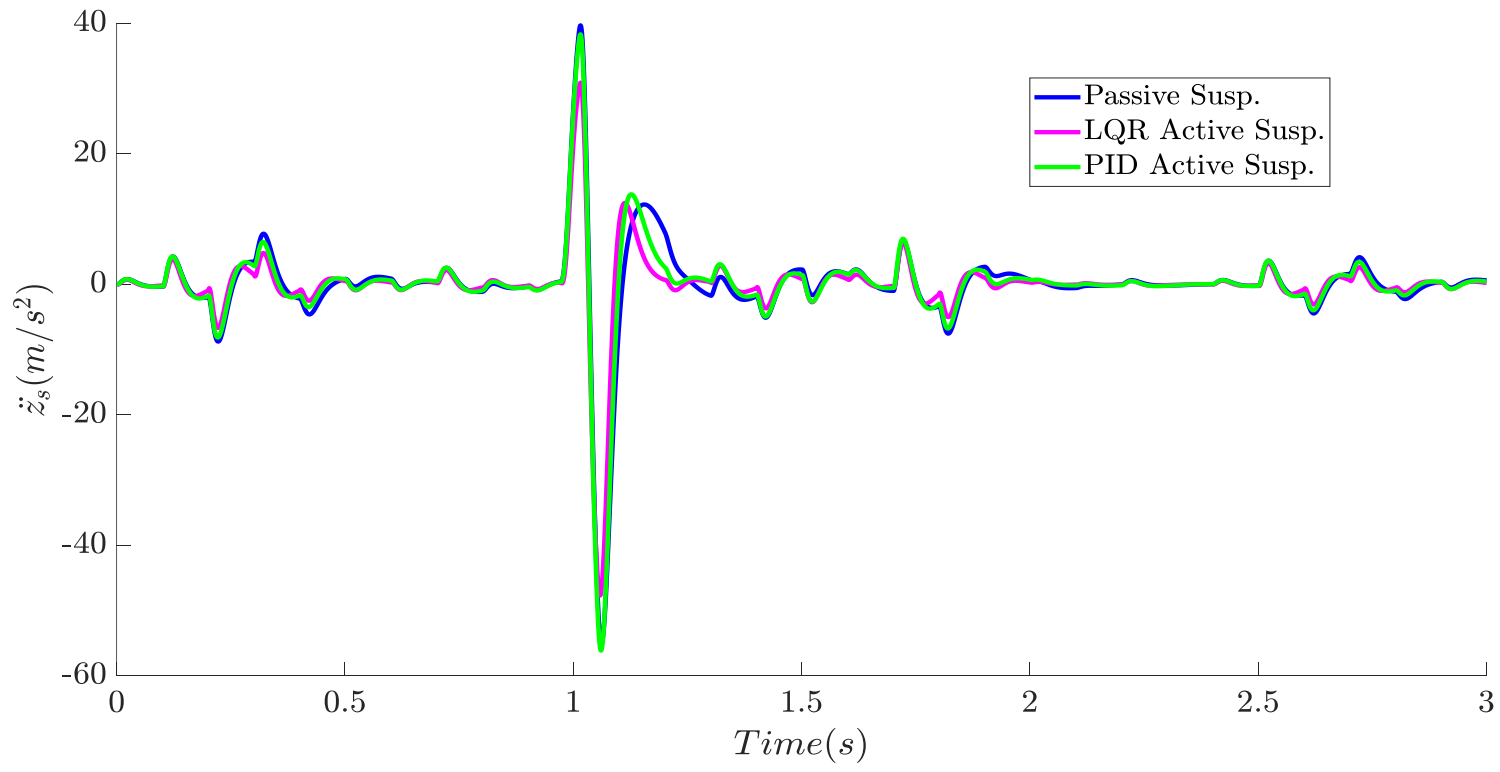

Fig. 18. Sprung mass acceleration under road disturbance with white noise

Table 10 shows the simulation of sprung mass acceleration under road disturbance with band-limited white noise.

**Table 10.** Time response for sprung mass acceleration under road disturbance with white noise

| Suspension type | Rise time (s) | Overshoot (m/s$^2$) | Settling time (s) |
|---|---|---|---|
| Passive | 1.835 s | -55.01 m/s$^2$ | 2.738 s |
| PID-controlled active | 1.733 s | -56.15 m/s$^2$ | 2.731 s |
| LQR-controlled active | 1.732 s | -47.71 m/s$^2$ | 2.727 s |

Examining the data in Table 10, it is clear that the active suspension with LQR control has the fastest rise time, the least overshoot, and the shortest settling time. Active suspension with a PID controller has the second-fastest rise time and the second-shortest settling time. Passive suspension has the longest rise time and the longest settling time. These results confirm that the active suspension with LQR control provides a more comfortable and safe ride than other suspensions.

Sprung mass motion simulation of suspension systems under road disturbance with band-limited white noise is presented in Fig. 19.

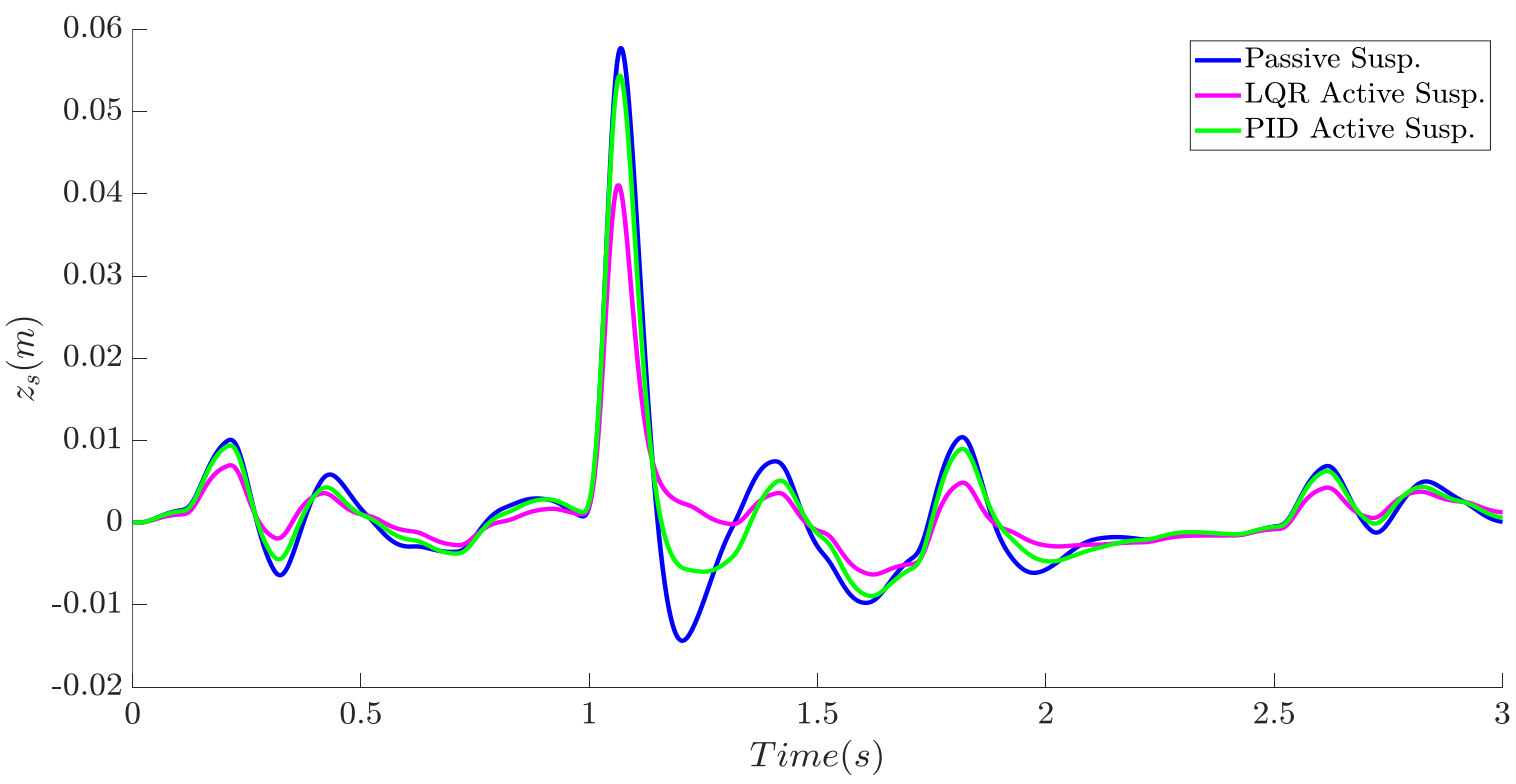

Fig. 19. Sprung mass motion under road disturbance with white noise

Table 11 presents the simulation of sprung mass acceleration under road disturbance with band-limited white noise.

**Table 11.** Time response for sprung mass motion under road disturbance with white noise

| Suspension type | Rise time (s) | Overshoot (m) | Settling time (s) |
|---|---|---|---|
| Passive | 1.270 s | 0.0577 m | 2.901 s |
| PID-controlled active | 1.202 s | 0.0543 m | 2.888 s |
| LQR-controlled active | 1.164 s | 0.0407 m | 2.872 s |

Analyzing the data in Table 11 reveals that the active suspension with LQR control has the fastest rise time, the least overshoot, and the shortest settling time. The active suspension with PID control is the second-best suspension in terms of these data. Passive suspension has the longest rise time, the most overshoot, and the shortest settling time. Thus, active suspension with an LQR controller has proven to be the most successful suspension.

## IV. Conclusion

In this research, LQR controller design and active suspension modeling were carried out for a quarter car. Comparative analysis was performed with passive suspension and active suspension with a PID controller to highlight the superiority of the LQR-controlled active suspension. Pole-zero maps of passive suspension, PID-controlled active suspension, and LQR-controlled active suspension were examined, and stability analysis was performed. As a result of the analysis, it was seen that the poles of all three suspension systems were to the left of the virtual axis and were stable. Suspension travel, sprung mass acceleration, and sprung mass motion simulations were performed under road disturbance, under simultaneous road disturbance and parameter uncertainty and under road disturbance with white noise for passive suspension, PID-controlled active suspension, and LQR-controlled active suspension. Rise time, overshoot, and settling time data of all three suspensions were obtained. When the obtained data were examined, it was seen that LQR-controlled active suspension was the suspension with the fastest rise time, least overshoot and the shortest settling time. In this case, it was proven that LQR-controlled active suspension was the most successful suspension system and provided a more comfortable and safe journey compared to passive suspension and PID-controlled active suspension.